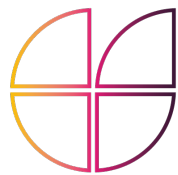
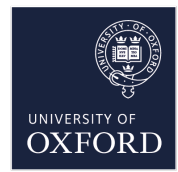

# Red Teaming Generative AI/NLP, the BB84 quantum cryptography protocol and the NIST-approved Quantum-Resistant Cryptographic Algorithms

Short title: Red Teaming Generative AI and Quantum Cryptography

Petar Radanliev[1], David De Roure[1], Omar Santos[2]

*Corresponding author email: radanliev@yahoo.com

[1]Department of Engineering Science, University of Oxford,

[2]Cisco Systems, RTP, North Carolina, United States

## Abstract

> In the contemporary digital age, Quantum Computing and Artificial Intelligence (AI) convergence is reshaping the cyber landscape, introducing both unprecedented opportunities and potential vulnerabilities.

This research, conducted over five years, delves into the cybersecurity implications of this convergence, with a particular focus on AI/Natural Language Processing (NLP) models and quantum cryptographic protocols, notably the BB84 method and specific NIST-approved algorithms. Utilising Python and C++ as primary computational tools, the study employs a "red teaming" approach, simulating potential cyber-attacks to assess the robustness of quantum security measures. Preliminary research over 12 months laid the groundwork, which this study seeks to expand upon, aiming to translate theoretical insights into actionable, real-world cybersecurity solutions. Located at the University of Oxford's technology precinct, the research benefits from state-of-the-art infrastructure and a rich collaborative environment. The study's overarching goal is to ensure that as the digital world transitions to quantum-enhanced operations, it remains resilient against AI-driven cyber threats. The research aims to foster a safer, quantum-ready digital future through iterative testing, feedback integration, and continuous improvement. The findings are intended for broad dissemination, ensuring that the knowledge benefits academia and the global community, emphasising the responsible and secure harnessing of quantum technology.

**Note about the lead author:**

Dr Petar Radanliev

Oxford e-Research Centre, Department of Engineering Science, University of Oxford

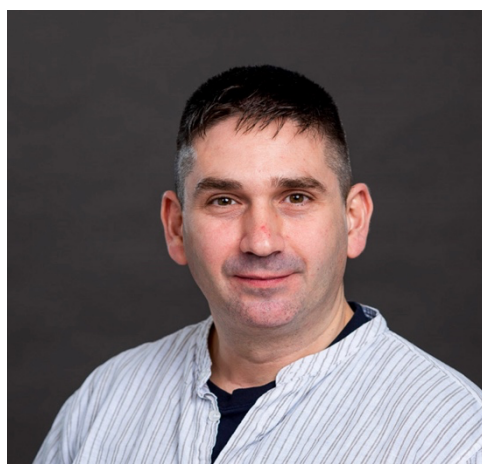

Petar Radanliev is a Post-Doctoral Research Associate at the University of Oxford. He obtained his PhD at the University of Wales in 2014. He continued with Postdoctoral research at Imperial College London, the University of Cambridge, Massachusetts Institute of Technology, and the University of Oxford. His current research focuses on Artificial Intelligence, the Internet of Things, Cybersecurity, Quantum Cryptography, and Blockchain Technology. Before joining academia, Dr Petar Radanliev worked as a Cybersecurity manager for the Royal Bank of Scotland for ten years, and before that, as a Lead Penetration tester for the Ministry for Defence for five years.

## 1. Introduction: Quantum Technology, AI, and the Evolving Cybersecurity Landscape

In the contemporary technological epoch, the rapid evolution of Quantum Computing and Artificial Intelligence (AI) is reshaping our digital realm, expanding the cyber risk horizon. As we stand on the cusp of a quantum revolution, the cyber-attack surface undergoes a transformation, heralding a future rife with potential cyber threats.

# University of Oxford

**Petar Radanliev, BA Hons., MSc., Ph.D.**
POSTDOCTORAL RESEARCH ASSOCIATE

## 1.1. Theoretical Underpinning

This research endeavours to construct a robust cybersecurity framework, ensuring AI's harmonious and secure integration with the Quantum Internet. Central to our exploration is evaluating AI/Natural Language Processing (NLP) models and their interaction with quintessential quantum security protocols, notably the BB84 method and select NIST-endorsed algorithms. Leveraging the computational prowess of Python and C++, we aim to critically assess the resilience of these quantum security paradigms by simulating AI-driven cyber-attacks.

## 1.2. Research Objectives

Envision a quantum-enhanced internet, operating at unparalleled speeds, yet fortified against AI-mediated cyber threats. This vision encapsulates our primary objective: to ensure that the digital advancements of the future, powered by AI, remain benevolent and secure. Over a five-year trajectory, our mission is to harness AI's potential in a manner that is beneficial and safeguarded against malevolent exploits. This research study is crafted with a primary endeavour to construct a formidable cybersecurity framework, aiming for seamless integration between AI and the Quantum Internet. Our focal point lies in the rigorous safety assessments of AI/NLP models and a comprehensive evaluation of quantum computing security protocols, notably the BB84 method and specific algorithms endorsed by NIST.

## 1.3. Methodological Approach

Our research methodology is rooted in comprehensive literature reviews, enabling a profound understanding of the current quantum communication landscape and the inherent AI risks. After this foundational phase, we use computational modelling, employing the BB84 protocol and NIST-sanctioned techniques. We introduce AI elements by integrating Python and C++, probing for potential vulnerabilities within these security frameworks.

A cornerstone of our research strategy is the "red teaming" approach. Emulating potential adversarial entities, we assess the robustness of quantum security protocols. Utilising AI models, enriched by datasets from esteemed sources such as Cornell ArXiv and Penn Treebank, we simulate cyber-attacks on these quantum defences to uncover and fortify any detected vulnerabilities.

By leveraging powerful programming languages like Python and C++, our research will gauge the resilience of quantum security mechanisms under AI-driven penetration attempts. We envisage a future where the internet, supercharged by quantum technology, operates at unparalleled speeds without compromising security. Our objective over a meticulous five-year research trajectory is to champion the cause of AI integrations that stand beneficial and intrinsically secure. By the culmination of this endeavour, our insights aim to act as a catalyst, galvanising further secure and ethically grounded research in this realm.

Our first step to methodically undertake this challenge involves an exhaustive literature review, gleaning insights into the present landscape of quantum communication and associated AI vulnerabilities. This foundational knowledge will formulate computational models tailored for rigorous testing. Python and C++, in synergy with the BB84 method and NIST-sanctioned algorithms, will host AI mechanisms geared to pinpoint potential security breaches.

The flowchart in Figure 1 provides a visual representation of the paper's structure, starting with the introduction and moving through various sections, including methodology, results, and conclusion. Each section is divided into sub-sections to provide a detailed overview of the paper's content.

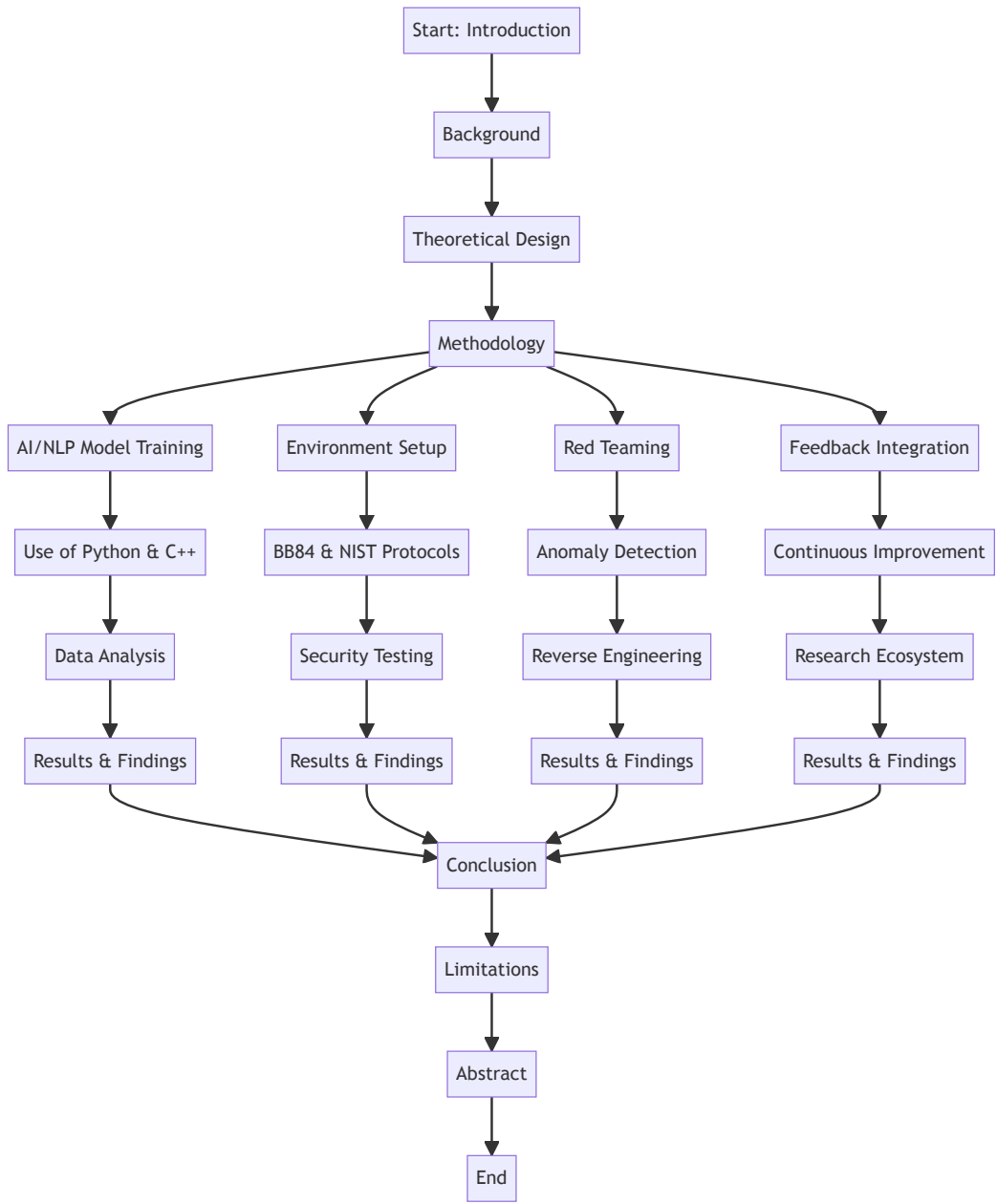

*Figure 1: Flowchart visually represents the paper's structure.*

An intrinsic component of our investigative approach remains 'red teaming'. This simulation of adversarial tactics aims to assess the potency and resilience of existing security measures. Leveraging AI models, enriched by esteemed datasets from sources such as Cornell ArXiv and Penn Treebank, challenges will be mounted against quantum security frameworks. Any identified vulnerabilities will undergo a detailed analytical process, culminating in remediation strategies and strengthening protocols.

## 1.4. Knowledge Dissemination and Broader Implications

Our commitment transcends the academic corridors of the University of Oxford. We envision a global knowledge-sharing ecosystem, disseminating our findings through diverse channels, from academic journals to public workshops. As we navigate the quantum future, the pertinence of our research will only amplify, laying the groundwork for a secure and responsible quantum era. Our overarching aspiration is to ensure that as quantum technology permeates industries, governments, and societies, its immense power is harnessed judiciously and securely.

# 2. Defining the terms and introducing the aims and objectives

## 2.1. Cryptography

Cryptography, translated from its original meaning in Greek, is 'secret writing'. Good cryptography depends on the hardness of the match problem. In other words, the algorithm is only as strong as the math problem of the specific cryptographic algorithm. Second is the quality of implementation, as correct implementation is fundamental for algorithm security. Third is the key secrecy because secret keys must be stored somewhere, usually by a centralised trusted authority. If you are a hacker trying to hack a cryptosystem, you will start with one of these three things: a hacker would try to solve the math problem, look for vulnerabilities in the implementation, or get access to the secret keys. Of the three risk scenarios, only the first is a cryptography problem. The second and third scenarios are more of an operational risk and cybersecurity problems.

The first risk scenario, the hard maths problems, is typically represented in the type of algorithm used, or in other words, as symmetric vs. public key encryption.

### 2.1.1. Symmetric

Symmetric key cryptography is when one key is used to encrypt and decrypt information, and the most well-known standard in this category is the Advanced Encryption Standard (AES), also known as Rijndael [1], based on the name of the creator Vincent Rijmen. Rijndael was selected to be the AES by the U.S. National Institute of Standards and Technology (NIST) in 2000 and has been an official standard since 2001 [2].

### 2.1.2. Asymmetric

Asymmetric cryptography is also known as public-key cryptography, uses two different keys, one is public key that is used for encryption and is known to all, and second is the private key that is used for decryption and is only known by one party. The most famous algorithm for public-key cryptography is the RSA cryptosystem developed in 1977 [3] Other well-known and frequently used algorithms include: the Digital

Signature Algorithm (DSA), which is based on the Schnorr and ElGamal signature schemes [4]; the Diffie–Hellman key exchange over public channels [5]; or as others have referred to as a method for 'secure communications over insecure channels' [6]; or the Elliptic-curve cryptography (ECC) that is based on algebraic structure of elliptic curves over finite fields. One point that is quite interesting to mention, while the RSA cryptosystem was publicly described the algorithm in 1977, the British mathematician and cryptographer Clifford Cocks, while working for the GCHQ in the year 1973, described an equivalent system in an internal document [7], what this brings to lights is that knowledge discovery is a process that follows a linear pattern. Hence, although we do not know how to develop or even implement new types of cryptography, the knowledge developed is ongoing, and the question is not whether the solutions will be developed, but who will develop the new solutions first.

## 2.2. Quantum cryptography

Quantum cryptography utilises specific physical laws to enhance the computational complexity of mathematical algorithms for securing information. Quantum cryptography exploits the so-called superposition of quantum particles. In a superposition, these particles, qubits, can pick any values between 0 and 1. For example, a 2-qubit computer thus can run operations with all four combinations (0-0, 0-1, 1-0, 1-1) of the quantum states simultaneously, potentially significantly surpassing the computation power of any current computers. Theoretically, such an improvement can eliminate the large computational and time costs of breaking a cryptographic algorithm.

Besides improving the computational speed, another attribute of quantum cryptography is that the very act of quantum particle observation changes its state. Any attempt to eavesdrop on a quantum communication will be detectable, as it would alter the transmitted particles' state. The security of the key exchange relies on the fact that any attempt to intercept or measure the qubits will disturb their quantum state.

Most current cryptography is not quantum-safe, meaning a powerful quantum computer can theoretically break the cryptographic keys. This led mathematicians to develop quantum-safe cryptographic algorithms. The most well-known quantum cryptography protocol, "quantum key distribution" (QKD), involves the transmission of a random sequence of quantum bits or "qubits" between two parties. These qubits, such as photons or electrons, can be encoded in various physical systems. The best-known QKD is the BB84 protocol published by Bennett and Brassard in 1984 [8]

IoT devices and other embedded systems with limited computational power can find it particularly challenging to generate strong cryptographic keys today. With quantum computers, their communication will be one of the most exposed. Therefore, NIST announced a special competition to develop quantum-safe cryptographic methods for these computationally weak devices [9].

## 2.3. Aims and objectives.

### 2.3.1. Objective 1

The future integration of Artificial Intelligence (AI) with the Quantum Internet has opened a novel frontier laden with both transformative prospects and profound security implications. Our research aims to pioneer the application of Penetration Testing (commonly known as Red Team exercises) to contemporary Generative AI and Natural Language Processing (NLP) models. This, in conjunction with quantum cryptographic procedures, will predominantly focus on the BB84 protocol [10] and the suite of NIST-endorsed Quantum-Resistant Cryptographic Algorithms [11], [12]. Harnessing Python and C++, our rigorous endeavour will initiate AI-driven penetration assays on quantum cryptographic architectures, emulate real-world quantum scenarios, and devise mitigation methodologies for identified vulnerabilities. The study outlines the penetration testing phases and the salient objectives that encompass identifying and rectifying frailties within the BB84 protocol, thereby refining its cryptographic resilience and culminating in developing a fortified quantum-secure prototype.

### 2.3.2. Objective 2 - Quantum cryptography vs post-quantum cryptography

*Post-Quantum Security*

When we have a large-scale quantum computer built, it will break all public-key cryptography widely used today [13]. This includes the tools (e.g., EC/EdDSA, VRFs, ZK proofs) used by all major blockchains. NIST is aware of this and has initiated the PQC process (2016-present). In this study, we consider two potential risk scenarios.

*Two risk scenarios from a future large-scale quantum computer*

**Risk scenario one** is a future large-scale quantum computer that can be used to attack the progress and development of protocol and hardware implementation. This means, it could forge transactions and send money, steal money from banks and crypto wallets. These cyber risks can be described as forward-looking attacks, but these are not the main concerns of the present, because we need to start worrying about these risks once there is a large-scale quantum computer, which might happen sometime in the future.

**Risk scenario two** is what we can describe as a 'time-travel' attack, which means that a future large-scale quantum computer can go back in time and rewrite history, it could forge medical records and replace existing patient records, it could cause a denial of service, or even replace (rewrite) the complete history (e.g., of bank digital records, blockchain records, medical records). The risk is that when a large-scale quantum computer is developed in 10-15 years from now, it can go back and look at the historical encrypted data records, break the cryptoscopy used today, and rewrite history to cause chaos in the present.

*Objective – to protect from future quantum attacks.*

This study aims to protect today's chain from future quantum attacks. To protect the protocol and network of the future, we need to protect today's digital assets. One method that we propose to test is 'State Poofs', which are digital certificates that can confirm that a sufficient total stake is verifiable, or non-verifiable.

*Solutions to test - cheap verification, outside of the network, SNARK-friendly*

The **solutions** that need to be investigated and tested include (1) cheap verification; (2) verification is done outside of the main network protocol; (3) is SNARK-friendly. The 'State Proofs' is a proof of state of a transaction or change in the digital asset and would operate like what we currently refer to as a dual factor authenticator. The 'State

# University of Oxford

**Petar Radanliev, BA Hons., MSc., Ph.D.**
POSTDOCTORAL RESEARCH ASSOCIATE

Poofs' solution is a smart contract that preserves the distributed security, where the smart contract would periodically be provided with 'State Poofs'. This solution would be combined with verifying one proof of state per 100s of rounds and enable verification of the network (or the blockchain) from the network's genesis, or the last good known state, enabling post-quantum security assurances. To ensure the safety of this system, the proof of state solution needs to be highly decentralised, and it must be infeasible to create a minority proofs. The second key concept of this solution is that we need a centralised authority to assemble and redistribute the proofs (the same as the dual factor app on your mobile). Still, this authority must be considered an untrusted provider and contain only a small and random fraction of the smart proofs. For this solution to be operational, the smart proofs must be converted into tiny and cheap SNARKs. This could be resolved with deterministic (e.g., Falcon [11]) verification that is mostly linear, and SNARK friendly.

*The benefits – expected outcomes*

The benefits of this solution are proofs of the state that can be implemented in the networking protocols and architecture for easy verification of the state by entities outside of the network. The solution is based on distributed quantum computing and adds long-term post-quantum security to the networking protocols and architecture. The implementation can be ultra-compressed into tiny and cheap SNARKs. This solution adds long-term post-quantum security because to create a proof of state in a distributed system, you need to have a certain contribution from the network, a fraction (around 70-80%) of the stake attested. Without that fraction, it is infeasible to create a stake proof, even if you had a quantum computer. In other words, if a quantum computer tries to create a fake proof of stake, the 'State Poofs' would confirm the previous state. The solution also improves network interoperability, because by converting the proof of state into a compressed SNARK (e.g., zk-SNARK proofs).

## 3. Research Methodology

### 3.1. Methodology to determine the importance of the study

Our study is underpinned by the seminal pronouncements made at the Black Hat and DEF CON 31 conferences (August 2023) by governmental dignitaries and titans of the tech world. At DEF CON 31, a congregation witnessing our active participation and logistical contribution in the Red Hat Village, the hacking cohort, and pivotal stakeholders in Generative AI discourse underwent a paradigmatic shift. On Black Hat's concluding day, the White House unexpectedly disclosed its collaborative venture with AI luminaries - including OpenAI, Google, Antrhopic, Hugging Face, Microsoft, Nvidia, and Stability AI, culminating in a public appraisal of generative AI ecosystems at DEF CON 31.

### 3.2. Methodology to determine the quality of the study.

The duality of ground-breaking evolution and intrinsic vulnerabilities lies in the nexus of AI and the Quantum Internet. Our postulate asserts that a specialised red team approach, amalgamating AI/NLP blueprints with the quantum cryptographic tenets of the BB84 protocol and NIST-sanctioned algorithms, unveils latent security lacunae, thereby fortifying quantum internet infrastructure [14]. Through the synergistic capabilities of C++ and Python, our investigation is poised for intricate depth and adaptability to surmount multifaceted quantum cryptographic enigmas. The methodological core is anchored in the versatile roles of Python and C++, exemplifying their composite prowess in achieving strategic orchestration and granular computational might.

### 3.3. Methodology to determine the potential impact.

The expanding acclaim of large language models (LLMs) like ChatGPT indicates a transformative phase in textual and communicative paradigms. Yet, these advancements come infused with intrinsic susceptibilities. Perils such as confabulations, jailbreak scenarios, and inherent biases are no longer confined to cybersecurity mavens but resonate with the collective societal conscience. In accord and conjunction with the White House's Office of Science, Technology, and Policy, we are poised to helm a research expedition dedicated to the forensic assessment of these emergent generative AI constructs. With the White House's explicit endorsement for such autonomous evaluative endeavours [15], we posit our methodology, rooted in Red Teaming paradigms, as a beacon aligning with the foundational principles of the Biden administration's AI Bill of Rights [16] and the AI Risk Management edicts decreed by the National Institute of Standards and Technology [17]. By definition, "red teaming" encapsulates a proactive security ethos, where specialists assume adversarial roles to challenge, evaluate, and enhance the defensive robustness of systems and frameworks.

### 3.4. Timeliness Given Current Trends, Context, and Needs

During DEF CON 31, the AI Village's founder accentuated a crucial challenge: the prevailing issues with Generative AI models remain unresolved owing to a knowledge gap in their red team evaluation. Building upon insights from the PETRAS project [18], our study develops the design for executing the UK's most comprehensive red team exercise on select AI models. Our study will differentiate from contemporaneous endeavours by targeting quantum cryptography, emphasising the BB84 protocol and NIST's Quantum-Resistant Cryptographic Algorithms. Instead of working in isolation, we'll synergise with our American affiliates, including Cisco Systems and multiple US Cybersecurity Agencies (CSAF, NTIA, CISA, and NIST, among others). Our research outcomes will be disseminated at premier events like RSA, Black Hat, and DEF CON. Beyond this, our project envisages fostering collaborative efforts among the UK, EU, USA, and the global research community, leveraging expertise in AI vulnerabilities management.

### 3.5. Impacts on World-leading Research, Society, Economy, or the Environment

The intricacies of securing Large Language Models (LLMs) became strikingly evident at DEF CON 31, where participants interacted with LLMs in a controlled environment. Results from this event are slated for a February 2024 release, providing ample time for companies to rectify identified vulnerabilities. The very deferment of these findings underscores their gravity. Given DEF CON's massive participation, our research, conducted under the stringent ethical and privacy standards of the University of

# University of Oxford

**Petar Radanliev, BA Hons., MSc., Ph.D.**
POSTDOCTORAL RESEARCH ASSOCIATE

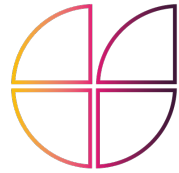 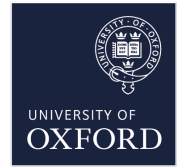

Oxford, offers a more secure avenue for assessing LLM vulnerabilities than a convention-based approach.

In Figure 2, the flowchart provides a visual representation of the research methodology, starting with the initial research proposal and moving through various stages, including theoretical design, background research, objectives definition, model training, environment setup, penetration testing, data collection, anomaly detection, reverse engineering, and feedback integration. The methodology concludes with the results and findings.

# University of Oxford

**Petar Radanliev, BA Hons., MSc., Ph.D.**
POSTDOCTORAL RESEARCH ASSOCIATE

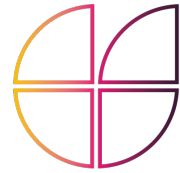
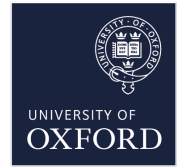

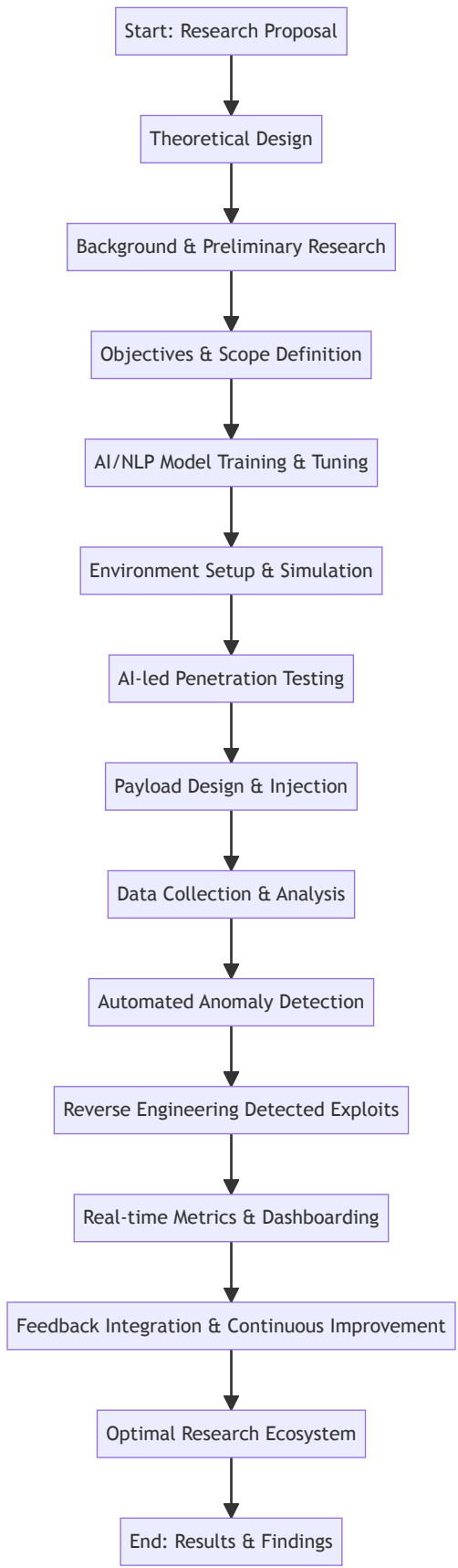

*Figure 2: Flowchart outlining the research methodology.*

## 4. Review of Novel AI and Quantum technologies and their significance

The new design for penetration testing of Generative AI and Quantum computing can produce several novel technologies in vulnerability management that could have wide-ranging impacts. Advanced AI/NLP models focused on vulnerability detection in cryptographic algorithms would be a significant step forward in cybersecurity. Quantum-resilient cryptographic protocols will be another focus, enhancing the security aspects of quantum computing. Automated penetration testing kits specifically designed for quantum systems will also be developed, representing a leap in the security evaluation process. Lastly, AI-driven platforms for efficient reverse engineering and enhanced payload delivery systems for quantum exploits could redefine how security assessments and countermeasures are developed. These technologies could be of particular interest to (Potential stakeholders) cybersecurity firms and agencies, academic researchers, and any industries where secure, fast communication is essential. Our proposed design is poised to unveil new vulnerabilities, leading to improved security of new technologies in the domain of vulnerability management, with potential reverberations across diverse sectors.

### 4.1. Red Teaming design

Identifying vulnerabilities in cryptographic systems is critical for secure communication in the digital age. This approach leverages Artificial Intelligence and Natural Language Processing (NLP) techniques to detect weaknesses in cryptographic algorithms. By combining these advanced technologies, we can significantly enhance the identification and mitigation of vulnerabilities. This represents a significant advancement in cybersecurity that safeguards against sophisticated cyber threats.

As the threat landscape evolves, it is crucial to take proactive measures to identify and address vulnerabilities. AI-driven methods have the potential to redefine cybersecurity standards, making systems more reliable and secure.

As quantum computing becomes more widespread, it is crucial to establish secure protocols that can withstand attacks using this advanced technology. Traditional cryptographic systems are vulnerable to quantum computers, making it necessary to develop quantum-resilient protocols to ensure safe communication in a post-quantum world. By enabling enhanced quantum security, we can instil confidence in the confidentiality and safety of quantum communications, leading to greater trust and adoption of this technology. The potential impact of this development is enormous, with the potential to revolutionise secure communication in the future.

As the usage of quantum systems continues to rise, it's imperative to ensure their security is up to par. Considering this, automated quantum pen-testing kits have been created to streamline evaluating their security. These advanced kits are engineered to automatically test the security of quantum systems, providing users with a comprehensive overview of their current security status. By harnessing the power of these kits, quantum systems can be fortified against potential security breaches, ultimately safeguarding their functionality.

Cutting-edge solutions have emerged to tackle cybersecurity challenges, harnessing the power of AI to optimise reverse engineering tasks and facilitate payload delivery systems that combat quantum exploits. Reverse engineering is crucial for identifying potential system weaknesses, and AI integration allows for greater precision and efficiency in detecting vulnerabilities. Advanced payload delivery systems ensure comprehensive security assessments addressing known and unknown threats.

These platforms can potentially revolutionise our approach to cybersecurity, leveraging AI capabilities to create more effective countermeasures and a safer digital environment. With AI-infused security assessment tools, we can optimise our security efforts and stay ahead of constantly evolving threats.

## 4.2. Ethical penetration testing

Our primary objective is to establish a strong and reliable framework for the upcoming quantum internet era. To achieve this, we focus on refining and enhancing the BB84 protocol, in conjunction with NIST-approved algorithms. We aim to ensure that all data transmissions remain secure and tamper-proof, which is crucial for building trust in digital communication.

Our security approach is unconventional as we look beyond traditional paradigms. We combine the power of Python and C++ to anticipate and neutralise potential threats before they arise. This forward-thinking strategy strengthens our security posture and establishes pre-emptive measures to safeguard systems from impending risks.

We strongly believe in the value of collaboration and knowledge sharing. Therefore, we are committed to sharing our findings and insights through scholarly publications and conference contributions. We aim to foster an environment of cooperation where shared knowledge is the driving force behind the development of quantum-safe innovations.

To prepare for the cybersecurity challenges that may arise with the advent of quantum computing, it is crucial to address current issues and anticipate potential future ones. Our main goal is to navigate this unexplored territory and lay the foundation for a future where the immense potential of quantum computing can be fully realised while minimising any risks that may arise.

Developing a moral framework beyond technological advancements is essential as AI continues to evolve. Our focus is on creating an AI landscape that prioritises ethical considerations. Considering these systems' increasing complexity and capabilities, our unwavering commitment is to ensure that their development and deployment always prioritise human safety and welfare.

Quantum communications involve more than just transmitting data; they encompass trust, privacy, and the many interactions connecting people worldwide. Our objective is to strengthen the quantum internet and usher in a new era of research and innovation. This period will be marked by a focus on security, ethical responsibility, and international cooperation. Through our efforts, we envision a future where UK institutions and the global community are prepared and eager to embrace the quantum era, with a strong emphasis on responsibility and security.

# University of Oxford
Petar Radanliev, BA Hons., MSc., Ph.D.
POSTDOCTORAL RESEARCH ASSOCIATE# 5. Red Teaming Design – Penetration Testing Design

## 5.1. Prototyping & Development

### 5.1.1. Refinement of Quantum Cryptographic Mechanisms

In our pursuit to enhance quantum cryptographic protocols, we have strategically harnessed the combined strengths of Python and C++. Our focus remains on the adaptation and elevation of the renowned BB84 protocol [10] and other NIST-endorsed quantum cryptographic methodologies, algorithms [11], [12], and cryptographic mechanisms [11], [19], [20]. Furthermore, we are venturing into the realm of AI integration, meticulously designing algorithms that seamlessly meld with our quantum frameworks [21]. These algorithms are poised to not only detect but also pre-emptively address vulnerabilities.

### 5.1.2. Simulation of Quantum-Secured Environments using AI/NLP Models

Our approach to model development is rooted in leveraging cutting-edge AI/NLP models. These models are designed to emulate potential adversaries and guardians within a quantum-secured milieu. Under the umbrella of Generative AI/NLP Integration, our objective is to employ Generative AI in simulating both conventional and malevolent user behaviours within a quantum network environment [22]. Foundational datasets like the Cornell ArXiv, supplemented by the Penn Treebank and WikiText, will serve as the bedrock for training our models in cryptographic contexts [23]. Our methodology is anchored in implementing avant-garde NLP techniques, with a specific emphasis on transformer-based models such as GPT variants [24]. The robustness and versatility of libraries like HuggingFace's Transformers will be pivotal in our NLP endeavours. To ensure the efficacy of our AI models, we will employ Python-based visualisation tools for continuous performance monitoring and analysis.

### 5.1.3. BB84 Quantum Cryptography Simulation

We aim to replicate the BB84 quantum key distribution protocol meticulously, facilitating AI interactions. Collaborations with esteemed entities like the Quantum Open-Source Foundation will be instrumental in procuring intricate data on BB84 implementations. C++, renowned for its computational prowess, will be the linchpin for the core computational facets of our quantum simulator. Potential integrations with platforms like Qiskit or QuTiP are on the horizon, ensuring swift and proficient simulation cycles. Python will be the cornerstone for scripting, automation, and data aggregation of retesting scenarios [25]. The culmination of these simulations will be a comprehensive documentation and analysis of AI's engagements with the BB84 protocol, spotlighting potential vulnerabilities.

### 5.1.4. Assessment of NIST Quantum-Resistant Cryptographic Algorithms

Our objective is clear: to rigorously assess the prowess of AI models against the NIST-sanctioned Quantum-Resistant Cryptographic Algorithms. The NIST database on Quantum-Resistant Cryptography will be our primary data source, offering insights into algorithmic implementations and their subsequent updates. Our methodology will encompass the integration of diverse algorithms, ranging from Lattice-based cryptographic methods to Code-based encryption techniques and Hash-based signatures. Python's expansive cryptographic libraries will be the backbone of our testing and automation processes. For tasks demanding heightened computational efficiency, C++ will be our go-to. The final evaluation will be a synthesis of Python and C++, aiming to benchmark the AI model's proficiency against quantum-resistant algorithms, identifying vulnerabilities, and fine-tuning performance.

### 5.2. Theoretical Framework for Real-world Quantum Network Testing: Field Testing and Validation

Quantum computing requires evaluating quantum systems' and protocols' real-world efficacy and vulnerabilities. This research constructs a theoretical framework for this purpose.

**Conceptual Foundations**:

1. **Quantum Network Dynamics**: Drawing from foundational principles of quantum mechanics and network theory, we postulate quantum networks' potential behaviours and challenges in real-world settings.
2. **User Interaction with Quantum Systems**: Grounded in human-computer interaction theories, we explore the nuances of end-user engagement with quantum systems, focusing on usability and potential user-triggered vulnerabilities.

**Data Sources and Methodological Considerations**:

1. **Collaborative Simulations**: By partnering with industry leaders, we aim to simulate authentic network scenarios, bridging the gap between theoretical postulations and practical applications.
2. **Synthetic Data Generation**: This approach, rooted in predictive modelling, seeks to emulate future quantum network behaviours, offering insights into prospective challenges and solutions.

**Proposed Theoretical Constructs**:

1. **AI/NLP-Driven Quantum Network Behaviours**: Integrating AI/NLP models with quantum simulations offers a novel perspective on network traffic behaviours, both typical and adversarial.
2. **User-Centric Quantum System Design**: By understanding end-user interactions and feedback, we can theorise optimal designs for quantum systems that are both secure and user-friendly.

**Evaluation and Knowledge Development**:

1. **Performance Metrics in Quantum Networks**: We can develop theories on optimal quantum network designs by identifying key indicators such as detection efficacy and system robustness.
2. **User Feedback Analysis**: A qualitative exploration of user feedback will contribute to the theoretical understanding of user needs, challenges, and potential system improvements in the quantum realm.

In Figure 3, we can visualise the emerging theoretical framework for penetration testing Generative AI and Quantum computers.

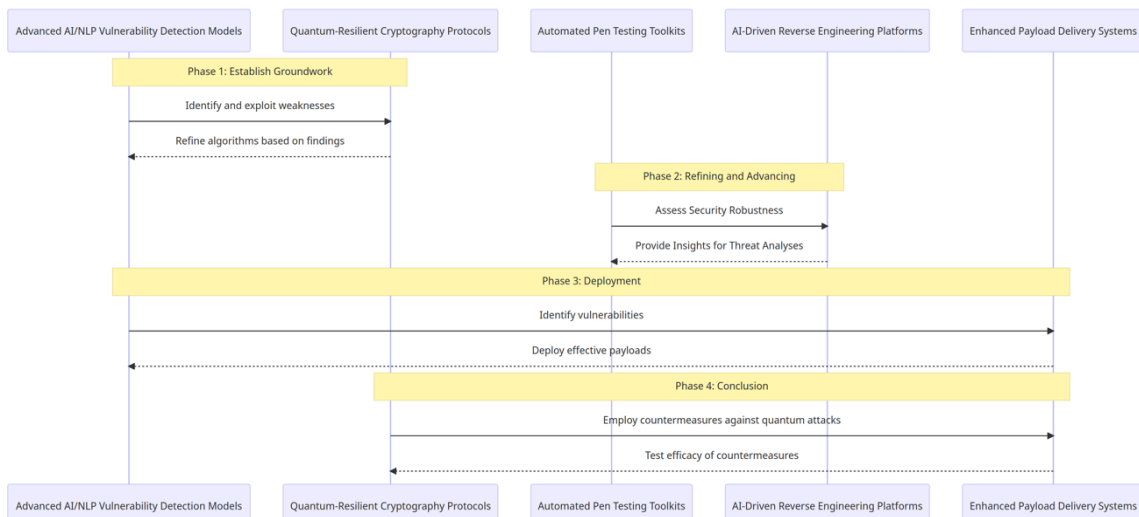

*Figure 3: Framework for penetration testing the convergence of Quantum Computing and Artificial Intelligence.*

In summary, our design is based on enhancing the theoretical comprehension of quantum networks in practical applications, focusing on bridging the gap between current knowledge and the vast potential of quantum computing. We will conduct rigorous evaluations and prioritise user-centric design considerations to significantly contribute to the quantum computing field.

## 5.3. Theoretical Framework for Post-Evaluation and Iterative Enhancement in Quantum-AI Systems

In the constantly evolving world of quantum computing and artificial intelligence, our methods require continuous analysis and adjustment. In this research project, we established a theoretical framework that will help improve quantum-AI systems' reviewing, optimising, and documentation processes. We can ensure these systems remain strong and practical in real-world applications by improving these crucial elements.

**Conceptual Foundations**:

1. **Iterative Quantum-AI System Design**: Drawing from iterative design principles, we postulate the significance of continuous refinement in quantum-AI systems, ensuring their adaptability and resilience.
2. **Documentation and Standardisation in Quantum Research**: Grounded in research documentation theories, we explore the importance of transparent, replicable, and standardised research practices in the quantum-AI domain.

**Data Sources and Methodological Considerations**:

1. **Feedback-Driven Data Collection**: By harnessing data from field testing, UAT feedback, and emerging research, we aim to create a comprehensive dataset that informs the iterative design process.

2. **Analytical Tools and Techniques**: Utilising Python's analytical capabilities and C++'s computational strengths, we propose a methodological approach to systematically identify and address areas of improvement.

**Proposed Theoretical Constructs**:

1. **Continuous Quantum-AI System Optimisation**: Integrating feedback and performance metrics, we theorise an Optimisation loop that ensures the evolution and relevance of quantum-AI systems.

2. **Research Documentation in Quantum Computing**: By collating research notes, datasets, and evaluations, we propose a structured approach to documenting quantum-AI research, ensuring its transparency, replicability, and relevance for future endeavours.

**Evaluation and Knowledge Development**:

1. **Performance Metrics in Iterative Design**: By comparing post-optimisation metrics against established benchmarks, we aim to develop theories on the effectiveness of iterative design in quantum-AI systems.

2. **Peer Review in Quantum Research Documentation**: A qualitative exploration of peer reviews will contribute to the theoretical understanding of research transparency, comprehensibility, and replicability in the quantum-AI domain.

Our theoretical framework for post-evaluation and iterative enhancement in quantum-AI systems attempts to enrich the theoretical understanding of post-evaluation and iterative refinement within quantum-AI systems. We aim to contribute substantially to the fundamental knowledge of quantum computing and AI by conducting thorough evaluation, meticulous documentation, and standardisation. These efforts set the foundation for future inquiries and the development of practical applications in this dynamic, interdisciplinary field.

## 5.4. Theoretical Framework for Collaborative Red Teaming in Quantum-AI Systems

In quantum computing and artificial intelligence, collaborative red teaming plays a crucial role. This section of our research endeavours to construct a comprehensive theoretical framework that elucidates how collaboration and feedback loops can enhance the potency and versatility of quantum AI systems.

**Conceptual Foundations**:

1. **Stakeholder-Centric Red Teaming**: Drawing from stakeholder theory, we postulate the significance of continuous engagement with key stakeholders in shaping and refining the red teaming process.

2. **Adaptive Threat Landscapes**: Grounded in adaptive systems theory, we explore the dynamics of threat environments that evolve in real-time, informed by AI/NLP feedback.

3. **Countermeasure Design and Iteration**: Leveraging iterative design principles, we delve into the processes of identifying vulnerabilities and crafting efficient countermeasures.

4. **Collaborative AI Learning**: Based on collaborative learning theories, we propose harnessing the collective intelligence of multiple AI models and expert insights to enhance threat simulation realism.

**Data Sources and Methodological Considerations**:

1. **Stakeholder Engagement Platforms**: We use communication platforms for virtual engagements and Python-based tools for collaborative data analysis to create a comprehensive feedback mechanism.
2. **Real-time AI/NLP Feedback Systems**: We envision a dynamic threat environment that mirrors advanced persistent threats by allowing AI models to adapt their strategies.

**Proposed Theoretical Constructs**:

1. **Feedback-Driven Red Teaming**: Integrating continuous stakeholder feedback, we theorise a red teaming approach that is both responsive and comprehensive.
2. **Adaptive AI Threat Simulations**: By allowing AI models to learn from their actions, we propose a threat simulation that evolves in real time, offering a more realistic representation of potential threats.
3. **Iterative Countermeasure Design**: Drawing from the identified vulnerabilities, we theorise an iterative approach to countermeasure design, ensuring maximum efficiency and adaptability.
4. **Ensemble Learning in Red Teaming**: By pooling knowledge from diverse AI models and expert insights, we propose a collaborative learning approach that enhances the realism and depth of threat simulations.

In this design development stage, we aimed to improve the understanding of collaborative red teaming in the context of quantum-AI systems. The design plan will achieve this through stakeholder engagement, adaptive threat simulations, iterative countermeasure design, and collaborative AI learning. By contributing to the foundational knowledge in quantum computing and AI, we hope to pave the way for future inquiries and practical applications in this interdisciplinary field.

5.5. Theoretical Framework for Quantum Network Behaviour Simulation and Refinement

As quantum computing advances, it's crucial to comprehend user actions, including malicious ones, in quantum networks. This study aims to create a theoretical framework that explains the objectives, scope, and methods for simulating these behaviours. The focus is on the BB84 protocol and NIST-endorsed Quantum-Resistant Cryptographic Algorithms.

Our conceptual foundations encompass three key areas: Quantum Network Behaviour Dynamics, AI/NLP Model Conditioning for Quantum Cryptography, and Emulated Quantum Environment Construction. Drawing from user behaviour theories and quantum mechanics, we analyse potential behaviours within quantum networks, distinguishing typical actions from malicious ones. Using machine learning theories, we also examine the conditioning of AI/NLP models to better understand quantum

# University of Oxford

**Petar Radanliev, BA Hons., MSc., Ph.D.**
POSTDOCTORAL RESEARCH ASSOCIATE

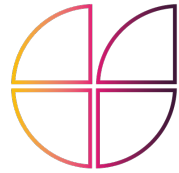 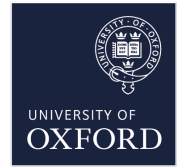

cryptography nuances and potential vulnerabilities. Finally, we explore creating a simulated quantum environment that mirrors real-world implementations of key quantum protocols and algorithms, leveraging simulation theories.

When defining objectives, working collaboratively with stakeholders is crucial to ensure alignment with real-world quantum computing challenges. To train the AI/NLP models, this framework proposes using foundational resources like Cornell ArXiv, Penn Treebank, and WikiText, which will prepare the scenarios for quantum cryptography simulations. Figure 4 shows a sequence diagram that includes steps, milestones, loops, critical points, and key outcomes.

# University of Oxford

**Petar Radanliev, BA Hons., MSc., Ph.D.**
POSTDOCTORAL RESEARCH ASSOCIATE

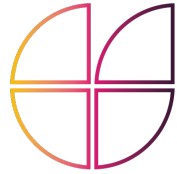
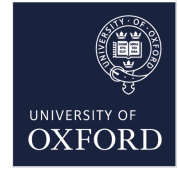

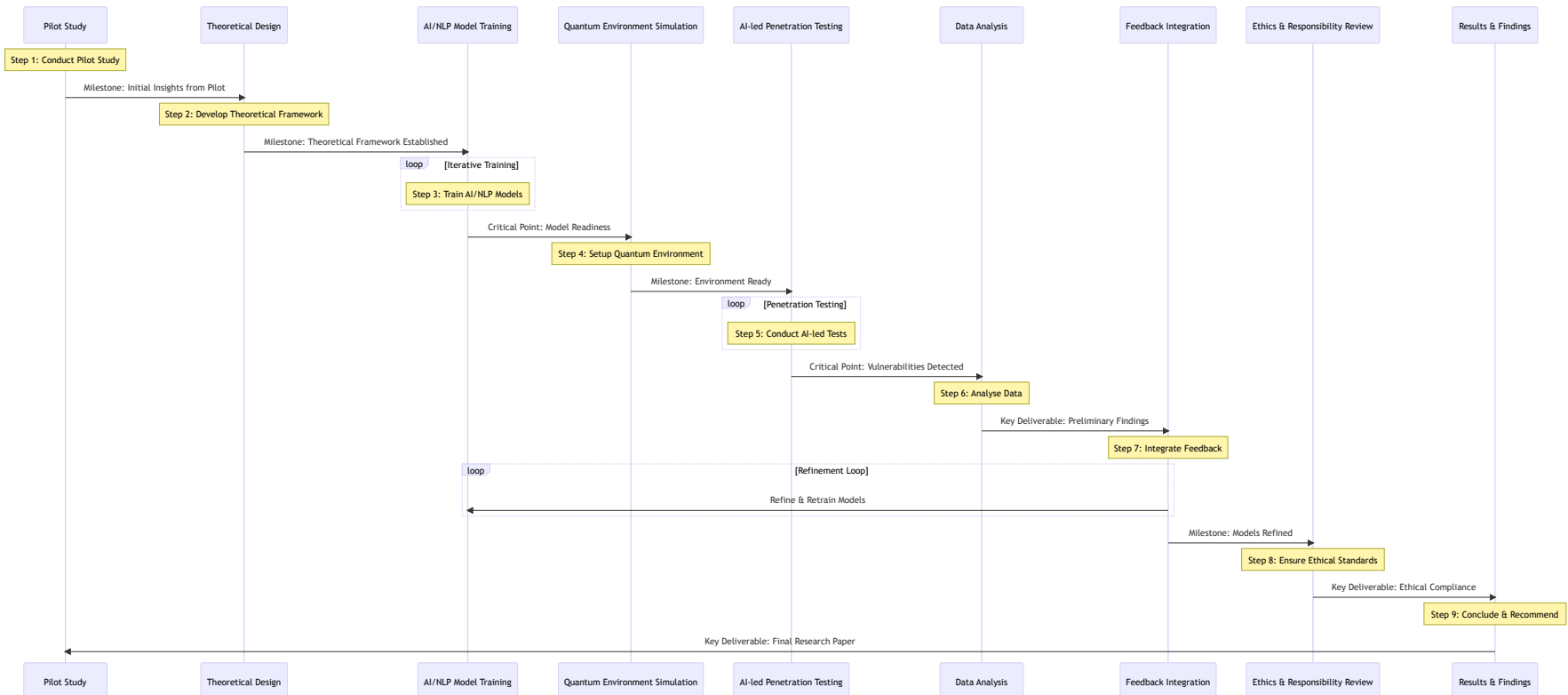

1. Figure 4: Sequence diagram of the steps, milestones, loops, critical points, and key outcomes from red teaming Generative AI and Quantum computers.

# University of Oxford

**Petar Radanliev, BA Hons., MSc., Ph.D.**
POSTDOCTORAL RESEARCH ASSOCIATE

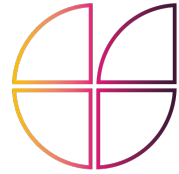 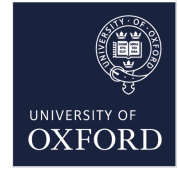

The sequence diagram in Figure 4 provides a more detailed representation of the emerging theory, highlighting steps, milestones, loops, critical points, and key deliverables throughout the research process.

# University of Oxford

**Petar Radanliev, BA Hons., MSc., Ph.D.**
POSTDOCTORAL RESEARCH ASSOCIATE

**Proposed Theoretical Constructs in** Figure 4:

1. **Environment Scanning and Validation**: By deploying Python scripts, we theorise an approach to scan and validate the quantum environment, ensuring its isolation and integrity.

2. **Efficient Quantum Simulation Development**: Utilising C++'s computational strengths, we propose the creation of a robust quantum simulation backbone, overlaid with Python's scripting capabilities for enhanced control and variability.

3. **Iterative Quantum Environment Optimisation**: Drawing from feedback loop theories, we postulate an iterative refinement approach for the quantum environment, leveraging AI insights to identify and rectify areas of enhancement.

We investigate the development and implementation of payloads that can effectively target and exploit the vulnerabilities in quantum systems by Utilising cybersecurity principles. Furthermore, our approach uses data analysis theories to gather, analyse, and present data derived from AI interactions, aiming to enhance the resilience of quantum communication systems. By integrating these cutting-edge technologies, we seek to advance our understanding of quantum communication and strengthen its security against potential threats.

When conducting penetration testing, there are various data sources and methodological considerations to remember. To deploy AI/NLP models, Python's dynamic capabilities are useful for managing model interactions and adjusting as needed. Meanwhile, C++ modules can help with swift exploit attempts. For developing and deploying payloads, C++ is particularly helpful due to its low-level control, while Python's automation capabilities come in handy for deployment sequences. Lastly, for in-depth data collection, Python's data analysis libraries, such as Pandas [26], can be useful for analysing AI interactions, with C++ being the go-to for high-frequency data capture needs.

Within the realm of quantum communication, we have developed a novel approach for proactively identifying and exploiting potential vulnerabilities in existing protocols. Our methodology, which we refer to as AI-Powered Quantum Protocol Interactions, relies on advanced AI/NLP models to enhance the security of quantum systems.

To ensure maximum adaptability and efficiency in the context of quantum exploits, we have incorporated a dual-layered payload design that integrates both C++ and Python. This design, which we call Optimised Quantum Payload Design, allows for greater flexibility and versatility in the face of evolving threat landscapes.

We take a continuous improvement approach to refine our quantum communication systems. By incorporating beta testing feedback and real-time data, we can continually enhance the resilience of our systems against potential threats. This iterative approach, which we call Continuous Improvement of Quantum Systems, is crucial for maintaining the highest system security and reliability levels.

Our framework is built upon three foundational concepts. Firstly, we recognise the critical role of real-time monitoring and alerting mechanisms in detecting deviations from established quantum system behaviours. Drawing from anomaly detection and quantum mechanics theories, this foundation is pivotal to our work. Secondly, we delve into methodologies for dissecting successful exploits to better understand their mechanics and develop countermeasures. This foundation builds upon cybersecurity and reverse engineering theories. Lastly, we explore the importance of providing real-time insights on red teaming activities, vulnerabilities, and countermeasure effectiveness using data visualisation theories.

**Data Sources and Methodological Considerations**:

1. **Continuous Quantum Environment Monitoring**: Python's statistical and ML capabilities will be harnessed to establish behavioural baselines and detect deviations, complemented by C++'s efficiency for real-time anomaly detection.
2. **Exploit Analysis Tools**: Tools like Radare2's r2pipe API, IDA Pro, and Ghidra will be instrumental in dissecting and understanding the intricacies of detected exploits.
3. **Dashboard Development for Real-time Insights**: Visualisation libraries like Matplotlib, Seaborn, or Dash will be pivotal in presenting key metrics and findings.

We have proposed several theoretical concepts that could benefit Quantum Red Teaming. The first concept entails implementing a system closely monitoring quantum environments, promptly flagging any unusual activity. This would enable us to identify potential threats before any damage is done.

Our second approach combines Python and C++ tools to reverse engineer potential exploits. Such a methodology would allow us to better understand how threats operate and how we can prevent them.

Lastly, we recommend employing data visualisation principles to create a real-time dashboard that offers clear insights into red teaming activities and system health. This would facilitate more effective system monitoring and ensure everything operates smoothly.

To improve red teaming strategies, we believe it's important to receive real-time feedback and adapt to newly identified threats in the constantly changing quantum communication landscape. To achieve this, we draw from feedback loop theories and quantum mechanics.

Our approach to iterative design principles involves reviewing findings post-testing, gathering stakeholder feedback, and planning subsequent testing iterations. By enhancing the quantum system in this way, we can continually improve our red teaming strategies.

To effectively communicate our findings, we leverage data visualisation and narrative theories. Interactive reporting tools like Jupyter Notebooks for an interactive report [27], allow us to weave together narratives, code snippets, and visuals to inform the review process.

**Data Sources and Methodological Considerations**:

1. **Post-Testing Review Mechanisms**: Python's data analysis and visualisation capabilities will be harnessed to dissect findings and inform the review process, complemented by interactive tools like Jupyter Notebooks.
2. **Rapid Implementation of Feedback**: C++'s efficiency will be pivotal in swiftly implementing changes to simulation environments and AI interaction routines, ensuring quantum systems remain resilient against evolving threats.

**Proposed Theoretical Constructs**:

1. **Feedback Integration in Quantum Red Teaming**: By continuously gathering and integrating feedback, we theorise an approach that ensures red teaming strategies remain updated and relevant, addressing the ever-evolving threat landscape.
2. **Interactive Quantum Red Teaming Reporting**: Utilising Python's Jupyter Notebooks, we propose a comprehensive reporting methodology that offers clear insights, narratives, and actionable findings from red teaming activities.
3. **Swift Quantum System Refinements**: Drawing from rapid development principles, we postulate an approach that swiftly implements feedback-driven changes, ensuring quantum systems' resilience against contemporary threats.

## 6. Discussion: Societal Benefits from Penetration Testing of Generative AI and Quantum Computing

In today's rapidly evolving technological landscape, society must stay ahead of the curve and ensure that the systems and solutions being developed are secure and advanced. One of the key areas of focus in this regard is quantum cryptography initiatives. By investing in research and development in this field, we can create solutions that are more resistant to attacks and provide more security for individuals and organisations.

Another important aspect of ensuring a safe digital environment is to examine the intricacies of the AI threat landscape, particularly within quantum frameworks. By understanding the potential vulnerabilities and risks associated with these technologies, we can better address and mitigate AI-driven cyber threats. This helps protect against attacks and promotes a safer and more secure online experience for everyone.

In addition to enhancing security, adopting AI for automated assessments can streamline processes and save valuable time. This efficiency can lead to more sustainable technological advancements that benefit a wider audience. By incorporating insights from research into technological products and solutions, we can solidify their position as cutting-edge tools that address modern-day challenges and requirements.

Furthermore, the refinement and potential licensing of protocols like BB84 can create new revenue streams, boosting the economy and allowing for reinvestment in research and development. This endeavour has the potential to bridge gaps between academia, technology giants, and cybersecurity experts, leading to holistic solutions, knowledge sharing, and a more interconnected and informed society. Investing in quantum cryptography initiatives and examining the AI threat landscape are crucial to ensuring a safe and advanced technological future for us all.

In the dynamic fields of quantum mechanics and cybersecurity, it is critical to consistently fine-tune cryptographic protocols in response to emerging vulnerabilities. Drawing from these domains, we delve into the refinement process.

To ensure the durability of these refined protocols, we rely on validation theories to examine them against AI/NLP models. This enables us to detect and mitigate potential threats. When collaborating on quantum cryptography initiatives involving AI resources, adhering to the highest ethical standards is imperative. We utilise ethical theories and AI principles to investigate the importance of this alignment.

We have proposed some new theoretical constructs for quantum cryptography. Our first approach involves identifying vulnerabilities and refining cryptographic protocols to address the ever-evolving threat landscape. We also suggest using AI/NLP models to validate the robustness of these protocols against potential threats. Lastly, we recommend adhering to the highest ethical standards, drawing on ethical principles, especially when working on quantum projects that involve AI tools.

At the heart of AI/NLP advancements lies the need for effectiveness and continuity. It is essential to establish a solid foundational framework in the early stages and continually refine and improve it over time. Using project management and AI theories, we emphasise this approach.

Feasibility and risk management are also crucial considerations in AI/NLP development. Incorporating regular assessment intervals and audits into the project timeline, we can detect and address challenges early on, as guided by risk management principles.

In the realm of cybersecurity, a methodological blueprint for AI/NLP development is essential. We explore the potential of merging automated and manual assessments to ensure the continued relevance of robustness toolkits developed internally. This approach draws on both cybersecurity and AI development theories.

We propose three theoretical constructs to enhance advancements in AI/NLP and cybersecurity projects. The first approach involves structured planning by assigning specific deliverables and milestones each year to ensure the project's direction remains aligned with its overarching goals. The second approach involves proactive risk management by Recognising potential challenges and implementing a comprehensive risk management strategy that includes recurrent audits, diverse funding avenues, and contingency frameworks. Finally, we suggest an iterative methodology for AI/NLP in cybersecurity that combines automated and manual assessments to ensure the continued relevance and robustness of AI/NLP models.

This section discusses the foundations of our framework and how we have built upon prior methodologies and findings. We believe it is important to extend our work over a longer period and introduce new models, milestones, and evaluations to ensure that our research insights remain relevant and adaptable to technological shifts and unforeseen challenges.

University of Oxford

**Petar Radanliev, BA Hons., MSc., Ph.D.**
POSTDOCTORAL RESEARCH ASSOCIATE

We explored the significance of translational outcomes in AI/NLP cybersecurity, leveraging translational research principles to transition our research outputs into academic papers, technological handovers, and practical applications. Our proposed theoretical constructs include extended research horizons in AI/NLP, refining earlier models and ushering in cutting-edge fortification phases. We will also discuss translational milestones in AI/NLP projects, emphasising the transformation of research into meaningful impacts through comprehensive strategies encompassing scholarly publications, technological advancements, and real-world applications.

Finally, we draw from collaborative research principles to postulate an approach that integrates collaborations with industrial stakeholders, ensuring that our research insights are channelled into practical solutions and policy inclusions for societal impacts in AI/NLP research.

We believe the key to optimal research is a strategic location at the intersection of academia, advanced research facilities and industry partnerships. This ecosystem can encourage unparalleled collaboration, ultimately enhancing the quality of research.

Furthermore, we understand the importance of state-of-the-art infrastructure, equipped with cutting-edge computational tools tailored for AI/NLP models, to ensure successful project execution.

We theorise that a strategic location and collaborative advantages can significantly increase the probability of framework success, and we propose a comprehensive strategy that emphasises the transformation of research through state-of-the-art infrastructure, ensuring robust testing and model development.

## 7. Conclusion: Towards a Secure Quantum-AI Future

Upon delving deeper into the intricate interplay between Quantum Computing, Artificial Intelligence (AI), and cybersecurity, we can glean some noteworthy insights. These three domains are inextricably linked, as the unprecedented computational power of quantum computing can greatly enhance AI capabilities, which in turn can be harnessed to bolster cybersecurity measures. The potential impact of this convergence cannot be overstated, as it can revolutionise various industries and transform the way we approach data analytics, threat detection, and risk mitigation. It is, therefore, imperative that we continue to explore the implications of this nexus and develop strategies to maximise its benefits while mitigating any potential risks.

**1. The Evolving Digital Landscape**: The onset of Quantum Computing and AI heralds a transformative era in technology. While these advancements promise unprecedented capabilities, they also introduce a novel array of cyber threats. Recognising this duality is paramount, prompting a proactive approach to cybersecurity.

**2. Theoretical Foundations**: Our research, grounded in robust theoretical underpinnings, sought to construct a resilient cybersecurity framework. This framework ensures the harmonious integration of AI with the Quantum Internet, safeguarding the digital realm against potential threats.

**3. Methodological Rigour**: Through a meticulous methodological approach, we employed computational modelling, red teaming, and iterative testing. Our use of renowned quantum security protocols and AI simulations provided invaluable insights into potential vulnerabilities and their subsequent fortifications.

**4. Collaborative Ecosystem**: Positioned within the University of Oxford's technology precinct, our research benefited from a nexus of academic excellence, state-of-the-art infrastructure, and industry collaborations. This confluence enriched our research quality, ensuring comprehensive insights and practical applicability.

**5. Knowledge Dissemination**: Our commitment to global knowledge sharing underscores the broader implications of our research. By disseminating our findings through diverse channels, we aim to foster a global community equipped with the knowledge to navigate the quantum future securely.

**6. Future Directions**: While our research provides a foundational understanding of the AI-Quantum cybersecurity landscape, it also illuminates avenues for future exploration. The continuous evolution of technology necessitates ongoing research, ensuring that security remains at its forefront as the digital realm advances.

As we stand on the precipice of a quantum-enhanced digital age, the confluence of Quantum Computing and AI presents challenges and opportunities. Our research underscores the imperative of proactive cybersecurity measures, ensuring that the immense power of quantum technology is harnessed judiciously and securely. As we look ahead, we fervently hope that this research serves as a beacon, guiding future endeavours towards a secure and responsible quantum AI future.

## 7.1. Limitations

While our research offers significant insights into the intersection of Quantum Computing, AI, and cybersecurity, it is essential to acknowledge its limitations:

1. **Scope of Study**: Our research primarily focused on the BB84 protocol and specific NIST-approved algorithms. The vast realm of quantum computing and AI encompasses other protocols and models that were not explored in this study.

2. **Data Limitations**: The AI models were trained using datasets like Cornell ArXiv and Penn Treebank. While these datasets are comprehensive, they may not capture the entirety of quantum cryptographic nuances or the evolving nature of AI-driven threats.

3. **Technological Constraints**: Our reliance on Python and C++ for simulations, while efficient, might not capture the intricacies or vulnerabilities present in other programming environments or real-world quantum systems.

4. **Red Teaming Limitations**: While our red teaming approach simulated potential hacker activities, real-world cyber threats can be more diverse, sophisticated, and unpredictable than those replicated in controlled environments.

5. **Generalisability**: The findings, while pertinent to the conditions and parameters of our study, might not be universally applicable across different quantum or AI configurations or in varied geopolitical or technological contexts.

6. **Temporal Limitations**: The rapid evolution of both quantum computing and AI means that our findings, though relevant now, may require periodic re-evaluation to remain current in the face of technological advancements.

Recognising these limitations is crucial for the interpretation of our findings and provides avenues for future research to build upon and address these gaps.

# University of Oxford

**Petar Radanliev, BA Hons., MSc., Ph.D.**
POSTDOCTORAL RESEARCH ASSOCIATE

# University of Oxford

**Petar Radanliev, BA Hons., MSc., Ph.D.**
POSTDOCTORAL RESEARCH ASSOCIATE

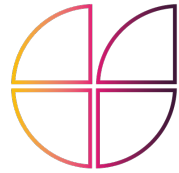
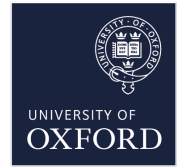